\documentclass[conference]{IEEEtran}
\IEEEoverridecommandlockouts

\usepackage[T1]{fontenc}
\usepackage{amsmath,amssymb,amsfonts}
\usepackage{algorithm}
\usepackage{algpseudocodex}
\usepackage{graphicx}
\usepackage{textcomp}
\usepackage[caption=false]{subfig}
\usepackage{float}
\usepackage{xcolor}
\usepackage{booktabs, multirow}
\usepackage{siunitx}
\usepackage{xspace}

\usepackage[colorlinks=true,allcolors=black]{hyperref}
\usepackage[numbers,sort&compress]{natbib}
\ExplSyntaxOn

\prop_new:N \g__jp_quantum_gates_prop
\prop_gset_from_keyval:Nn \g__jp_quantum_gates_prop
  {
    C3SX    = C\sp 3\sqrt{X}  , % 3-Controlled Sqrt(X)
    CCX     = CCX             , % Toffoli
    CCZ     = CCZ             , % Double Controlled Pauli Z
    CH      = CH              , % Controlled Hadamard
    CP      = CPhase          , % Controlled Phase
    CRx     = R\sb x          , % Controlled Rotation-X
    CRy     = R\sb y          , % Controlled Rotation-Y
    CRz     = R\sb z          , % Controlled Rotation-Z
    CS      = CS              , % Controlled Phase S
    CSdg    = CS\sp\dag       , % Controlled Inverse Phase S
    CSWAP   = CSWAP           , % Controlled SWAP
    CSX     = C\sqrt{X}       , % Controlled Sqrt(X)
    CU      = CU              , % Uniformly Controlled (2Q)
    CX      = CX              , % Controlled Pauli X (CNot)
    CY      = Y               , % Controlled Pauli Y
    CZ      = Z               , % Controlled Pauli Z
    DCX     = DCX             , % Double Controlled Pauli X
    ECR     = ECR             , % Echoed Cross-Resonance
    H       = H               , % Hadamard
    I       = I               , % Identity
    iSWAP   = iSWAP           , % iSwap
    P       = P               , % Phase Shift
    R       = R               , % Generic Rotation
    RC3X    = RC\sp 3X        , % 3-Controlled Margolus
    RCCX    = RCCX            , % Margolus (Simplified Toffoli)
    Rx      = R\sb x          , % Rotation X
    Rxx     = R\sb{xx}        , % Rotation XX
    Rxx-yy  = R\sb{xx-yy}     , % Rotation XX-YY
    Rxx+yy  = R\sb{xx+yy}     , % Rotation XX+YY
    Ry      = R\sb y          , % Rotation Y
    Ryy     = R\sb{yy}        , % Rotation YY
    Rz      = R\sb z          , % Rotation Z
    Rzx     = R\sb{zx}        , % Rotation ZX
    Rzz     = R\sb{zz}        , % Rotation ZZ
    S       = S               , % Phase S
    Sdg     = S\sp\dag        , % Inverse Phase S
    SWAP    = SWAP            , % Swap
    SX      = \sqrt{X}        , % Sqrt(X)
    SXdg    = \sqrt{X}\sp\dag , % Inverse Sqrt(X)
    T       = T               , % T Gate
    Tdg     = T\sp\dag        , % Inverse T Gate
    U       = U               , % Uniform Euler Rotation
    X       = X               , % Pauli X
    Y       = Y               , % Pauli Y
    Z       = Z               , % Pauli Z
  }

\NewDocumentCommand\gate{m}{
    \group_begin:
    \prop_get:NnNTF
      \g__jp_quantum_gates_prop
      { #1 }
      \l_tmpa_tl
      { \mode_if_math:TF
          { \mathtt{\tl_use:N \l_tmpa_tl} }
          { $ \mathtt{\tl_use:N \l_tmpa_tl} $ } }
      { \textcolor{red}{???} }
    \group_end:
}

\NewDocumentCommand\gates{sO{,~}m}{
    \group_begin:
    \IfBooleanTF{#1}{}{\texttt{\{}}
    \seq_set_split:Nnn \l_tmpa_seq , { #3 }
    \int_if_zero:nTF { \seq_count:N \l_tmpa_seq }
      { \textbf{\textcolor{red}{??}} }
      { \seq_pop:NN \l_tmpa_seq \l_tmpa_tl
        \exp_args:Ne \gate{ \tl_use:N \l_tmpa_tl }
        \seq_map_inline:Nn \l_tmpa_seq
          { #2 \gate{##1} } }
    \IfBooleanTF{#1}{}{\texttt{\}}}
    \group_end:
}

\ExplSyntaxOff

\title{Not All Qubits are Utilized Equally}
\author{%
  \IEEEauthorblockN{Jeremie Pope}
  \IEEEauthorblockA{\textit{School of Computer Science and Engineering} \\
    \textit{Pennsylvania State University}\\
    Centre County, PA, USA \\
    j.pope@psu.edu}
  \and
  \IEEEauthorblockN{Dr. Swaroop Ghosh}
  \IEEEauthorblockA{\textit{School of Computer Science and Engineering} \\
    \textit{Pennsylvania State University}\\
    Centre County, PA, USA \\
    szg212@psu.edu}
}

\begin{document}
\maketitle

\begin{abstract}
  Improvements to the functionality of modern Noisy Intermediate-Scale Quantum (NISQ) computers have coincided with an increase in the total number of physical qubits.
  Quantum programmers do not commonly design circuits that directly utilize these qubits; instead, they rely on various software suites to algorithmically transpile the circuit into one compatible with a target machine's architecture.
  For connectivity-constrained superconducting architectures in particular, the chosen syntheses, layout, and routing algorithms used to transpile a circuit drastically change the average utilization patterns of physical qubits.
  In this paper, we analyze average qubit utilization of a quantum hardware as a means to identify how various transpiler configurations change utilization patterns.
  We present the preliminary results of this analysis using IBM's $27$-qubit Falcon R4 architecture on the Qiskit platform for a subset of qubits, gate distributions, and optimization configurations.
  We found a persistent bias towards trivial mapping, which can be addressed through increased optimization provided that the overall utilization of an architecture remains below a certain threshold. As a result, some qubits are overused whereas other remain underused. The implication of our study are many-fold namely, (a) potential reduction in calibration overhead by focusing on overused qubits, (b) refining optimization, mapping and routing algorithms to maximize the hardware utilization and (c) pricing underused qubits at low rate to motivate their usage and improve hardware throughput (applicable in multi-tenant environments).
\end{abstract}

\begin{IEEEkeywords}
  quantum computing, quantum software engineering, transpilers, qubit utilization
\end{IEEEkeywords}

\section{Introduction}

The capabilities of quantum computers continue to grow \cite{2025/Buyya-Gill,2024/Sood-Chauhan,2022/Coccia} with significant improvements to the number and fidelity of physical qubits \cite{2025/AbuGhanem,2023/Castelvecchi} poised to advance us toward more fault-tolerant quantum hardware.
For chemistry \cite{2024/Weidman,2022/Sajjan,2022/Motta}, physics \cite{2022/Sajjan,2021/Andreas}, biomedical \cite{2022/Blunt,2021/Andreas}, and numerous other fields that stand to benefit from quantum computation, this transition is a prerequisite to realizing a practical ``quantum advantage'' over classical algorithms \cite{2023/Herrmann,2023/IBM/Davis,2023/Kim,2022/Beverland,2022/Proctor}.
With signs of this transition materializing through greater access to larger machines and the development of more efficient error correcting codes, it is impractical to design circuits for specific hardware architectures.
Instead, quantum programmers regularly rely on suites of algorithms to transpile circuits into a functionally similar form that is compatible with the target architecture's constraints \cite{2025/Murillo,2024/Sarkar,2021/Piattini}.

This process conceptually mirrors the use of classical compilers by abstracting away the need to handle resource allocation ourselves.
The "circuit mapping problem" is provably NP-hard \cite{2023/Ito,2023/Wagner,2015/Yamanaka}, prohibiting an exhaustive search for any but the most trivial circuits and architectures.
Through the combination of selected layout, routing, and optimization algorithms, the transpiler can offer an approximately optimal solution that, when configured correctly, is sufficient for most use cases.
A byproduct of this approach is that the mix of stochastic and heuristic-based methods used to navigate the search space inevitably introduces structural biases as to which physical qubits get allocated to the logical qubits.

Without an actual multi-tenant quantum computer, there has not been a need from a programmer's standpoint to ensure uniform resource allocation at the physical qubit level.
As a result, this indirect control over qubit placement remains largely uninvestigated.
From an architectural standpoint, however, understanding this relation can provide insight into the functional utility of a quantum computer outside the context of a particular workload.
By identifying "hot" and "cold" spots within the average qubit utilization qualified as a result of the transpiler configuration, we can more readily develop algorithms that utilize such biases, improve error through selective pruning of high-error, low-utilization qubits, or even serve as the grounds for monetary discounts as a means to improve the overall utilization of machines within a quantum-as-a-service execution model.

In this paper, we provide a novel method for measuring the influence various transpiler configurations have on average qubit utilization.
We discuss in-progress applications of this method using Qiskit's \cite{2024/Qiskit} transpilation pipeline, showing how configuration changes can influence placement within IBM's 27-qubit Falcon R4 architecture.

% - Pathfinder
The rest of this paper is structured as follows:
Section~\ref{sec:background} details information related to synthetic circuit generation and Qiskit's transpiler pipeline.
Section~\ref{sec:methodology} describes our methodology, outlining how parameters are captured alongside steps taken to isolate individual parameter's influence on
steps taken to increase reproducibility.
Section~\ref{sec:evaluation} presents our experimental setup, followed by analysis of results.
The limitations of the current analysis are also described.
We finally summarize key findings and provide closing remarks in Section~\ref{sec:conclusion}.

\section{Qiskit as a Quantum SDK}
\label{sec:background}

In this section, we describe how we generate a synthetic workload and then detail the staging of the Qiskit transpiler pipeline.

\subsection{Circuit Generation}
\label{sec:background:circuit-generation}

To test the transpiler, we generate synthetic workloads with a finite set of gates using Qiskit's \texttt{random\_circuit} function.
This function generates a seed-controlled random quantum circuit with a set number of qubits, target depth, and target gate-operand ratio.
For this analysis, the gate-operand ratio is always between one and two qubit gates, as the gates Qiskit supports with three or more operands are small.

Generation begins by first creating a list of gates that roughly matches the requested gate-operand ratio.
One qubit gates from the set \gates{H,I,P,R,Rx,Ry,Rz,S,Sdg,SX,SXdg,T,Tdg,U,X,Y,Z} and two qubit gates from the set \gates{CH,CP,CRx,CRy,CRz,CS,CSdg,CSX,CU,CX,CY,CZ,DCX,ECR,Rxx,Rxx-yy,Rxx+yy,Ryy,Rzx,Rzz,SWAP,iSWAP} are combined until the resulting list has the specified ratio of one to two qubit gates.
A blank circuit is created with the requested number of qubits.
The requested depth is generated by iteratively choosing values from the constructed gate list and randomly assigning the required qubit operands to the qubits in the circuit.
At the end of each iteration, qubits that did not get a gate have slack added to maintain alignment for the next cycle.

\subsection{Transpilation Pipeline}
\label{sec:background:transpilation-pipeline}

Qiskit divides transpilation into six distinct stages: initialization, layout, routing, translation, optimization, and scheduling.
These stages are detailed below, except for the scheduling phase, which Qiskit skips by default.

\subsubsection{Initialization}
This stage primarily serves as an extensibility interface for algorithms written for (but not distributed with) Qiskit.
By default, it decomposes multi-qubit gates into an equivalent set that only operates on two or fewer.

\subsubsection{Layout}
\label{sec:background:transpilation-pipeline:layout}

The layout stage provides the initial mapping of the virtual qubits of the quantum circuit to the physical qubits of the target architecture.

Unless the optimization level is greater than $1$, Qiskit first attempts a trivial mapping by assigning all virtual qubits the physical qubit with a matching index.
If this fails, or we start with a higher optimization level, the VF2++ algorithm \cite{2018/Juttner} runs to find an initial layout.
This algorithm treats the circuit as a connectivity graph and attempts to find a permutation that is a subgraph of the target architecture's connectivity graph.
The solution with the lowest permutation is selected, with ties decided through a noise minimization heuristic.
Qiskit achieves this using one of two methods: a density maximizing algorithm and the SABRE algorithm \cite{2024/Zou,2019/Li}.
When using SABRE, both layout and routing co-occur, effectively skipping the routing stage since the selected layout is already compatible with the target architecture's connectivity.

\subsubsection{Routing}
After selecting a layout, the circuit has all virtual qubits mapped to the target architecture's physical qubits, but might violate connectivity constraints.
The routing stage "rewires" the circuit by adding SWAPs that move gates operating over disconnected qubits to nearby connected pairs.
Doing this optimally is NP-Hard \cite{2023/Ito,2023/Wagner,2015/Yamanaka}, so Qiskit uses the SABRE algorithm's SWAP optimizer by default when using another layout method.
As a part of this process, SABRE can add previously unallocated physical qubits to the circuit if they provide a cost-minimizing path.
While these added qubits are categorically ancillary, Qiskit allocates all physical qubits to the circuit at the end of routing, regardless, so the distinction is lost.

\subsubsection{Translation}
Translation converts a circuit containing arbitrary quantum gates into one containing gates exclusively from the architecture's basis gate set.
This expansion, also known as unrolling, can significantly increase the gate count and depth of a circuit, as most modern architectures only support a limited number of native basis gates.

\subsubsection{Optimization}
Once fully expanded, optimizations deflate the circuit via substitution, cancellation, and synthesis, among other methods.
Qiskit controls this stage primarily through a coarse optimization level, ranging from \texttt{0} (no optimization) to \texttt{3} (heavy optimization).

In rare cases, a qubit ``optimizes out'' and becomes ancillary.
Since Qiskit always allocates all physical qubits, this is permitted, as the qubit is still treated as if it is in use.
The opposite is not true, however, as previously unutilized gates cannot gain gates without potentially violating a connectivity constraint.
If gates are added, the routing stage must be re-run, effectively treating this pass as happening both before and after routing.

Once all optimizer passes have completed, the circuit is entirely constructed and compatible with the target architecture.

\section{Methodology}
\label{sec:methodology}

In this section, we outline the process of parameter combination in a replicable and comparable manner. We also discuss how the test runner operates to compute the relevant statistics of a single circuit/transpiler configuration pair.

\subsection{Parameter Combination and Expansion}
\label{sec:methodology:parameter-expansion}

For reproducibility, we precalculate all combinations of generation and transpilation parameters before execution.
This pre-expansion enables the deterministic generation of configuration parameters, even when executing in parallel.

\begin{algorithm}[!t]
  \caption{Seed-Unique Parameter Expansion}
  \label{alg:parameters-expansion}
  \begin{algorithmic}[1]
    \ForAll{$(A_B,A_C,A_{\Delta T},A_S) \in \mathbf{A}$}
    \State Set global seed to $A_S$
    \State $A \gets \texttt{transpiler\_args}(A_B,A_C,A_{\Delta T})$
    \State $\mathbf{P} \gets \emptyset$
    \ForAll{$(q,d,r),(O,L) \in \prod (\mathbf{G},\mathbf{T})$}
    \State $G \gets \texttt{generator\_args}(q,d,r)$
    \State $T \gets \texttt{transpiler\_args}(O,L)$
    \ForAll{$\{i \in \mathbb{I} \mid 1 \leq i \leq M_G\times M_T\}$}
    \State $G_S \gets \texttt{random\_seed}(\mathbf{G})$
    \State $T_S \gets \texttt{random\_seed}(\mathbf{T})$
    \State $\mathbf{P} \gets \mathbf{P} \cup \{G,G_S,T,T_S\}$
    \EndFor
    \EndFor
    \ForAll{$(G,T) \in \mathbf{P}$}
    \State $\texttt{calculate}(A,G,T)$
    \EndFor
    \EndFor
  \end{algorithmic}
\end{algorithm}

Expansion occurs in two phases, as shown in \autoref{alg:parameters-expansion}.
For the first phase, we collect all architecture-specific information ($\mathbf{A}$), including the set of basis gates ($A_B$), qubit connectivity ($A_C$), minimum instruction time step ($A_{\Delta T}$), along with the global configuration seed ($A_S$).
We use the first three parameters separately over Qiskit's provided \texttt{BackendV2} object to filter any unexpected platform-specific optimizations, latent calibration data, or other settings that might change the behavior of the transpiler.
This is a limitation of the current implementation of this metric, which is discussed further in \ref{sec:limitations:no-calibration-data}.

After setting the global seed, the second phase computes the cartesian product of all non-seed circuit generation ($\mathbf{G}$) and transpilation ($\mathbf{T}$) configurations.
Excluding seed parameters from the initial product ensures that all combinations of non-seed parameters are paired with unique seeds across the entire parameter set.
With both the circuit generation and transpilation requiring seeds, this step avoids biasing collected data with the effects of replicated circuit or transpiler behaviors.

For the non-seed circuit generation parameters, we have the number of qubits used by the circuit ($q$), the target depth of the circuit ($d$), and the target ratio of one-qubit gates to two-qubit gates ($r$).
For the non-seed circuit transpilation parameters, we have the coarse optimization level ($O$) and layout algorithm ($L$).
The transpiler's routing algorithm is also investigated, but proved to scale poorly for options other than the default.

Repetition of the combined $\mathbf{G}$ and $\mathbf{T}$ parameters is controlled through $M_G$ and $M_T$, respectively.
Using these values, the generation seed $s_G$ and transpilation seed $s_T$ are generated in pairs.
The complete parameter set is then added to the parameter superset ($\mathbf{P}$) for use later.

Only after all configurations have been expanded can circuit evaluation begin.
We limit parameter expansion to a per-architecture basis to prevent clobbering of the global seed.

\subsection{Circuit Evaluation}
\label{sec:methodology:circuit-evaluation}

\begin{algorithm}[!t]
  \caption{Calculate Active Qubits}
  \label{alg:calc-active-qubits}
  \begin{algorithmic}[1]
    \Function{calculate}{$A,G,T$}
    \State $\mathbf{Q} \gets \emptyset$
    \State ${circuit}_1 \gets \texttt{generate}(G)$
    \State ${circuit}_2 \gets \texttt{transpile}(A,T,{circuit}_1)$
    \ForAll{$instr \in {circuit_2}$}
    \ForAll{$qubit \in instr$}
    \State $index \gets \texttt{indexof}(circuit_2, qubit)$
    \If{$index \notin \mathbf{Q}$}
    \State $\mathbf{Q} \gets \mathbf{Q} \cup index$
    \EndIf
    \EndFor
    \EndFor
    \State \Return $\mathbf{Q}$
    \EndFunction
  \end{algorithmic}
\end{algorithm}

The evaluation of a singular configuration triplet, shown above in \autoref{alg:calc-active-qubits}, begins with begins by generating an initial circuit (${circuit}_1$) using the method described in \ref{sec:background:circuit-generation}.
Next, we use Qiskit's \texttt{generate\_preset\_pass\_manager} function to convert the raw transpilation configuration into an executable pipeline.
Because we always provide $O$ as a part of $\mathbf{T}$, we explicitly control the coarse optimization level, which otherwise would default to $2$.

The constructed pipeline executed with $\mathbf{A}$, $\mathbf{T}$, and ${circuit}_1$ to generate ${circuit}_2$.
This process can be computationally intensive when performed over thousands of $(A,G,T)$ triplets.
Therefore, Qiskit defaults to leveraging multiprocessing where applicable.
With this function itself parallelized, we always explicitly turn off this feature by passing $num\_processes=1$ alongside the other parameters.

After transpilation, Qiskit considers all qubits in ${circuit}_2$ to be used.
To determine which qubits are used by actual instructions, we first iterate through the circuit, extracting each instruction ($instr$) before iterating over the instruction for each qubit.
Qiskit requires this double-iteration due to the way it exposes qubit indices between the internal C/C++ code and the external Python bindings.
A method on ${circuit}_2$ translates the qubit into a plain index, which is added to the set of active qubits ($\mathbf{Q}$) if unique.
After processing all qubits for all instructions, the resulting $\mathbf{Q}$ contains all active physical qubit indices.

To optimize memory pressure when executing this function in parallel, we transfer the parameter superset $\mathbf{P}$ to each execution process in its entirety prior to the first calculation.
We then convert the triplet into a set of indices in $\mathbf{P}$, which is reconstructed at the start of the function and back into the appropriate triplet.
The extra overhead this process requires significantly reduces the raw size of P while also avoiding computation and transfer overhead from pickling and unpickling potentially large Python structures, making it highly performant.

\section{Evaluation}
\label{sec:evaluation}

In this section, we briefly discuss the rationale for the selected parameters before continuing to observations collected from the in-progress data.

\subsection{Parameter Selection}

For this analysis, we consider the behavior of Qiskit's default compiler chain for IBM's 27-qubit Falcon R4 architecture under the \texttt{algiers} configuration.
All data is collected as described in \ref{sec:methodology:circuit-evaluation} using a parallel worker queue.

\subsubsection{Circuit Generation Parameters}

These parameters include $q={6,11,16}$, $d={20,40}$, and $r={4:1,1:11:4}$.
The limiting factor for this analysis was the number of qubits within the target architecture.
The value for $q$ was initially set to a value close to $20\%$ of the total number of qubits available, without going below.
The limit of $16$, representing just under $60\%$ of physical qubits, was chosen after observation found higher values produce similarly shaped results.
The depth $d$ was chosen after analysis showed little effect on observed patterns.
These two levels were explicitly chosen to provide a stark doubling of circuit depth.
Values for $r$ were chosen to provide a minimal layout between circuits that are populated by heavily $1$-qubit gates, equally distributed gates, and heavily $2$-qubit gates.

\subsubsection{Transpiler Parameters}

Transpiler parameters include only $O={1,2}$ and $L={trivial,dense,sabre}$.
The routing algorithm was not parameterized for this analysis as Qiskit only supports \texttt{basic}, \texttt{lookahead}, and \texttt{sabre} by default.
The former two values were excluded due to performance considerations as the architecture's supported number of qubits grows, leaving \texttt{sabre} as the only viable algorithm.
The values for $O$ were selected to toggle some, but not all, of the built-in optimizers.
$O3$ is specifically excluded because its behavior is similar to that of $O2$ across the other parameter combinations tested.
Values for $L$ include all default Qiskit layout algorithms, though the behavior of \texttt{trivial} and \texttt{dense} is similar given the selected routing algorithm.

\subsection{Experimental Results}

\subsubsection{Trivial and Dense Layout Methods}

\begin{figure}[!t]
  \centering
  \subfloat[$L=\mathtt{trivial}$]{%
    \includegraphics[width=0.8\linewidth]{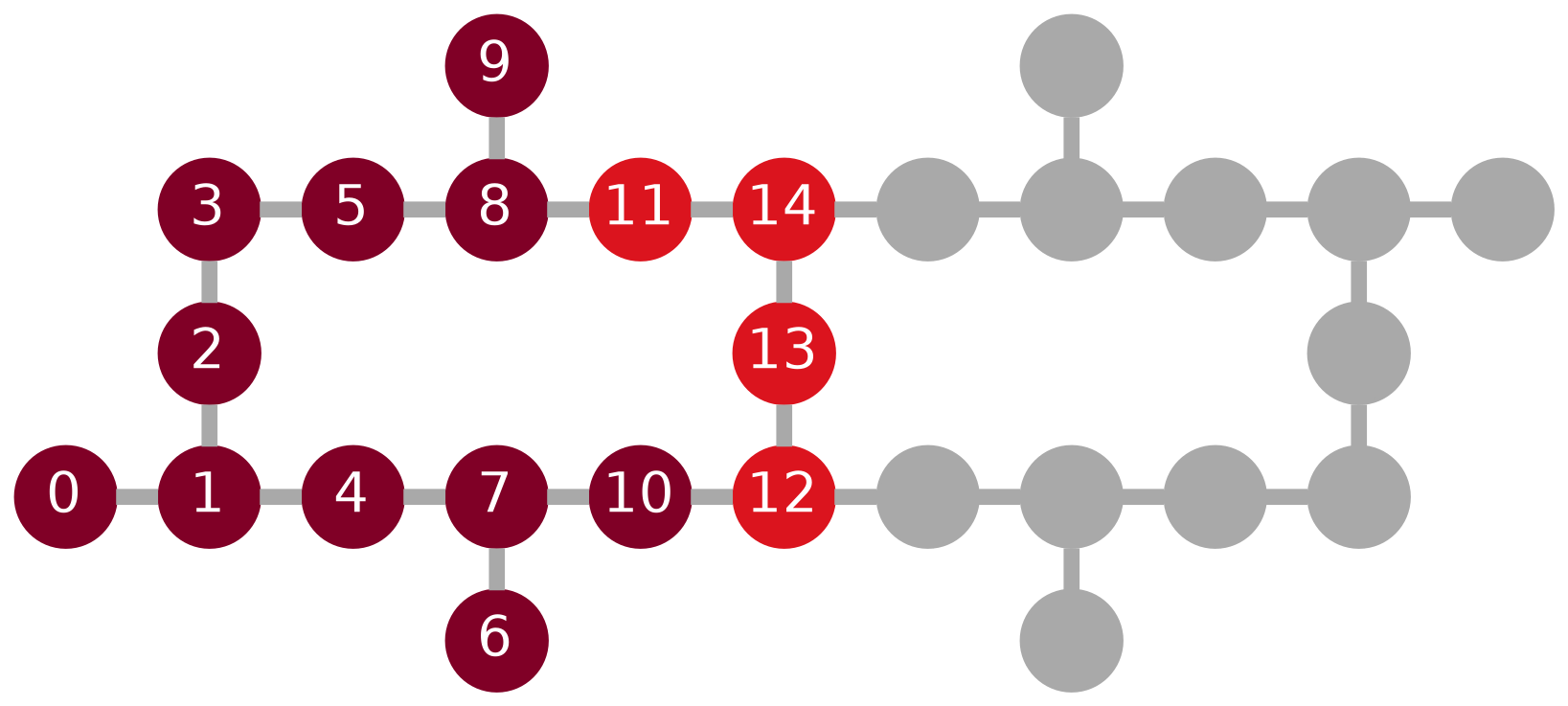}}
  \linebreak
  \subfloat[$L=\mathtt{dense}$]{%
    \includegraphics[width=0.8\linewidth]{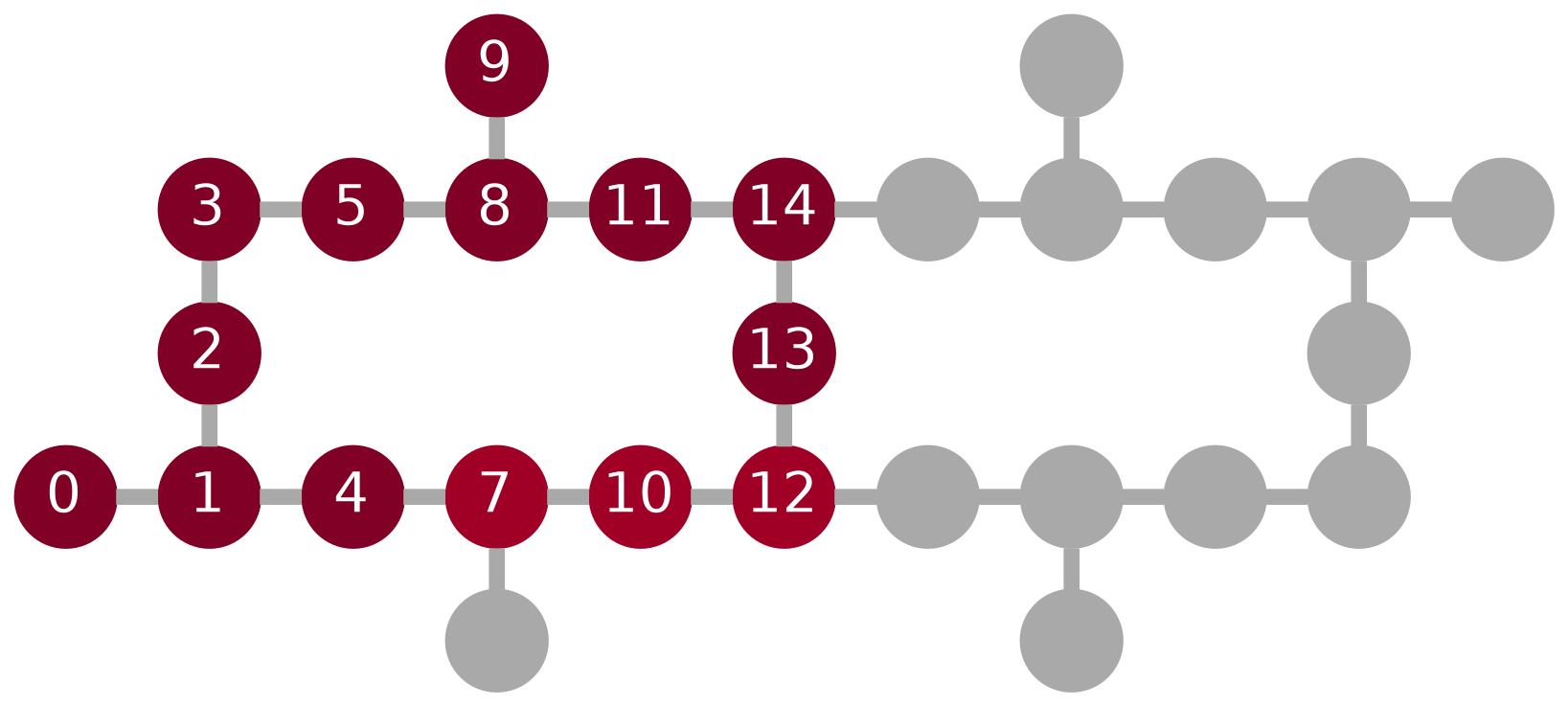}}
  \linebreak
  \includegraphics[width=0.8\linewidth]{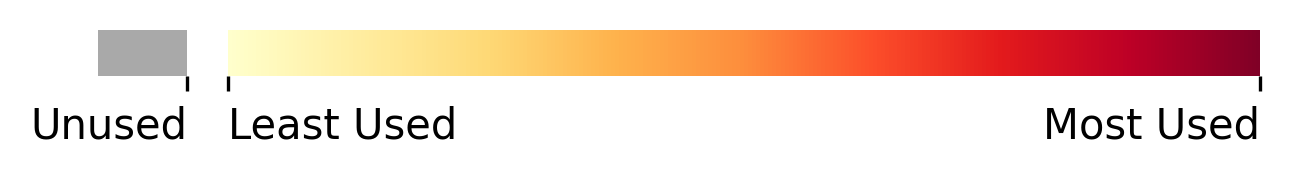}
  \caption{Average qubit utilization for $L=\mathtt{trivial}$ and $L=\mathtt{dense}$ with $q=11$, $d=20$, $r=\texttt{4:1}$, and $O2$ at a sample size of $N=400$. Shows how both algorithms produce similarly shaped results through different methods. \texttt{trivial} always uses indexes $0$ to $10$ which coincidentally aligns with \texttt{dense}, favoring highly connected qubits.}
  \label{fig:trivial-v-dense}
\end{figure}

Both the \texttt{trivial} and \texttt{dense} layouts exhibit similar behavior across the tested parameters.
This is not surprising as both methods use a naive approach to searching for and adding ancilla qubits.
The \texttt{trivial} layout linearly maps each qubit from the circuit onto the target architecture, starting at index $0$.
The \texttt{dense} layout attempts to maximize density by selecting the subgraph that has the highest connectivity.
As shown in \autoref{fig:trivial-v-dense}, both methods are incidentally similar for the tested parameters as they share the same routing method.
This leads to the \texttt{trivial} method always using indices $0-10$ but only occasionally requiring the additional $11$-$14$.
Similarly, the \texttt{dense} method always selects indices $0-5$, $8-9$, $11$, and $13-14$, as it produces the densest subgraph for $11$ qubits, incidentally including $7$, $10$, and $12$ is the most optimal route.

The overall meaningfulness of these two algorithms diminish as the overall architectural qubit count rises, so more complex analysis of how other configurations affect their behavior is not warranted.

\subsubsection{SABRE Layout vs. Optimization}

\begin{figure*}[!t]
  \centering
  \begin{minipage}{0.88\linewidth}
    \centering
    \subfloat[$r=\texttt{4:1}$ and $O1$]{%
      \includegraphics[width=0.32\linewidth]{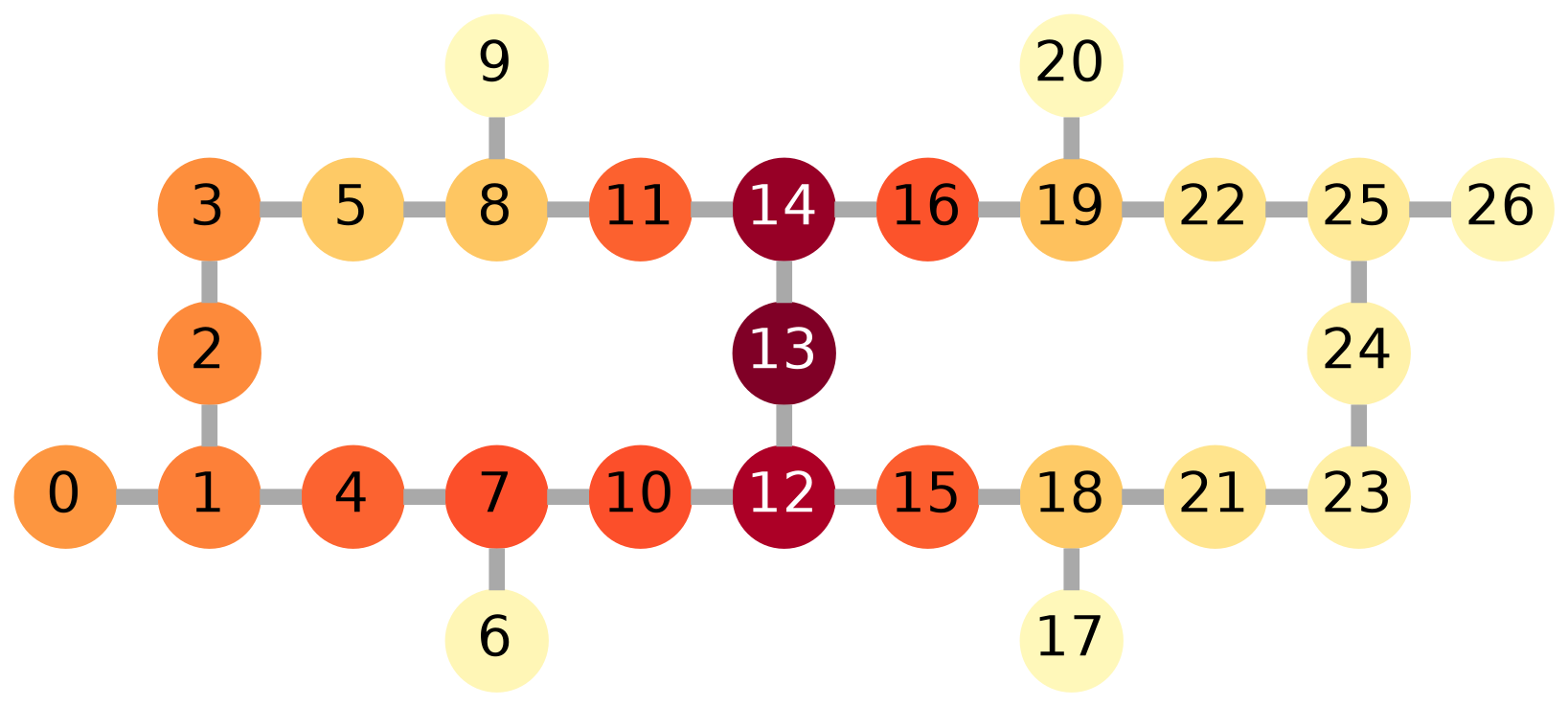}}
    \subfloat[$r=\texttt{1:1}$ and $O1$\label{fig:sabre:q6:o1:d20}]{%
      \includegraphics[width=0.32\linewidth]{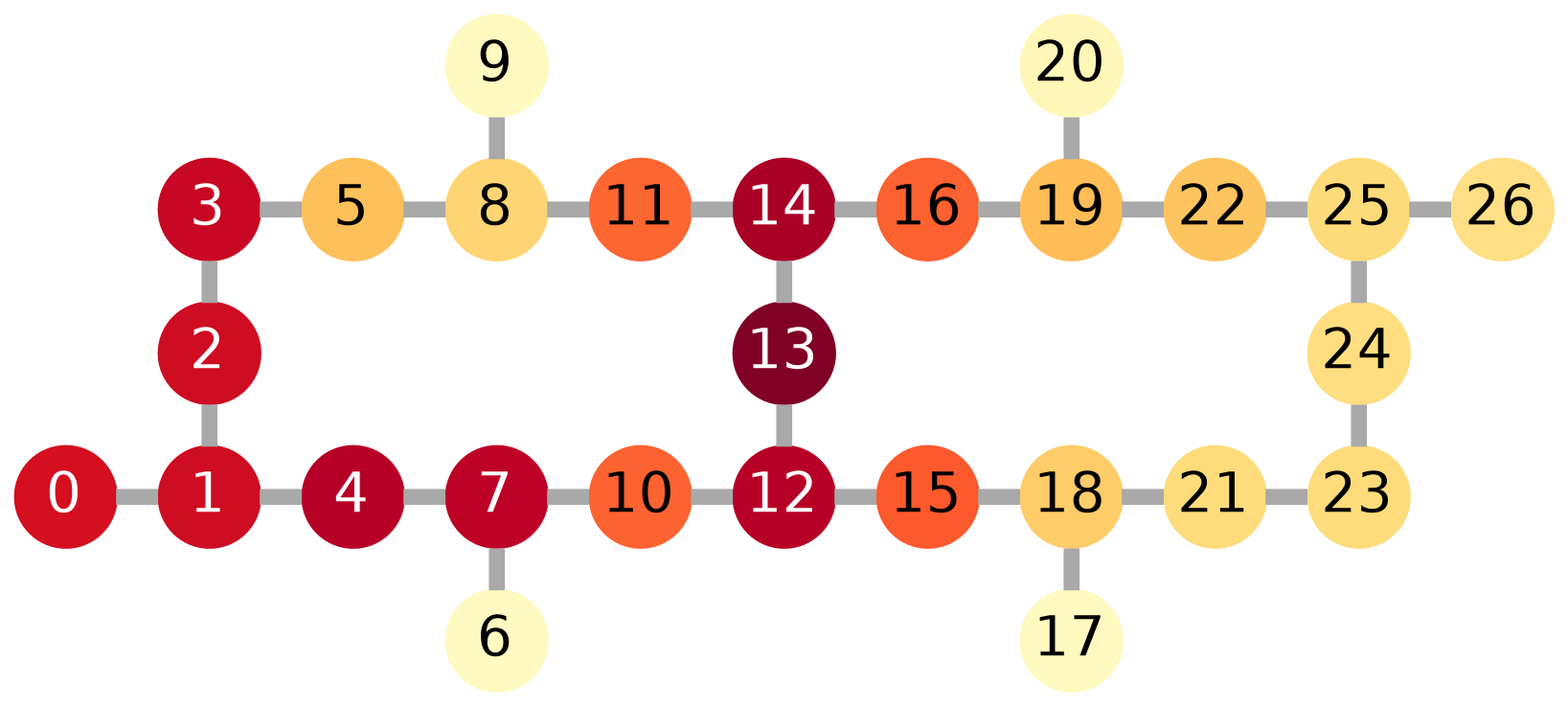}}
    \subfloat[$r=\texttt{1:4}$ and $O1$]{%
      \includegraphics[width=0.32\linewidth]{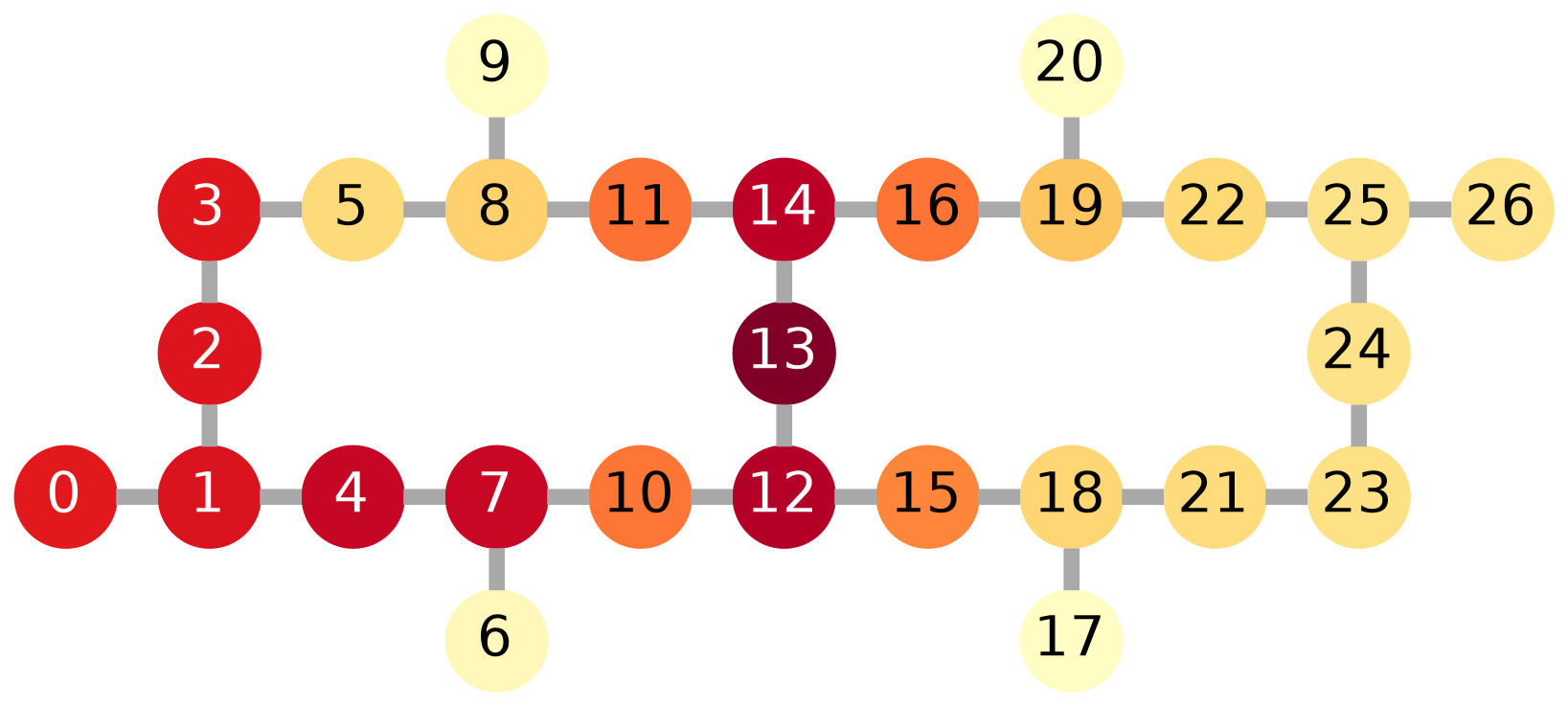}}
    \linebreak
    \subfloat[$r=\texttt{4:1}$ and $O2$]{%
      \includegraphics[width=0.32\linewidth]{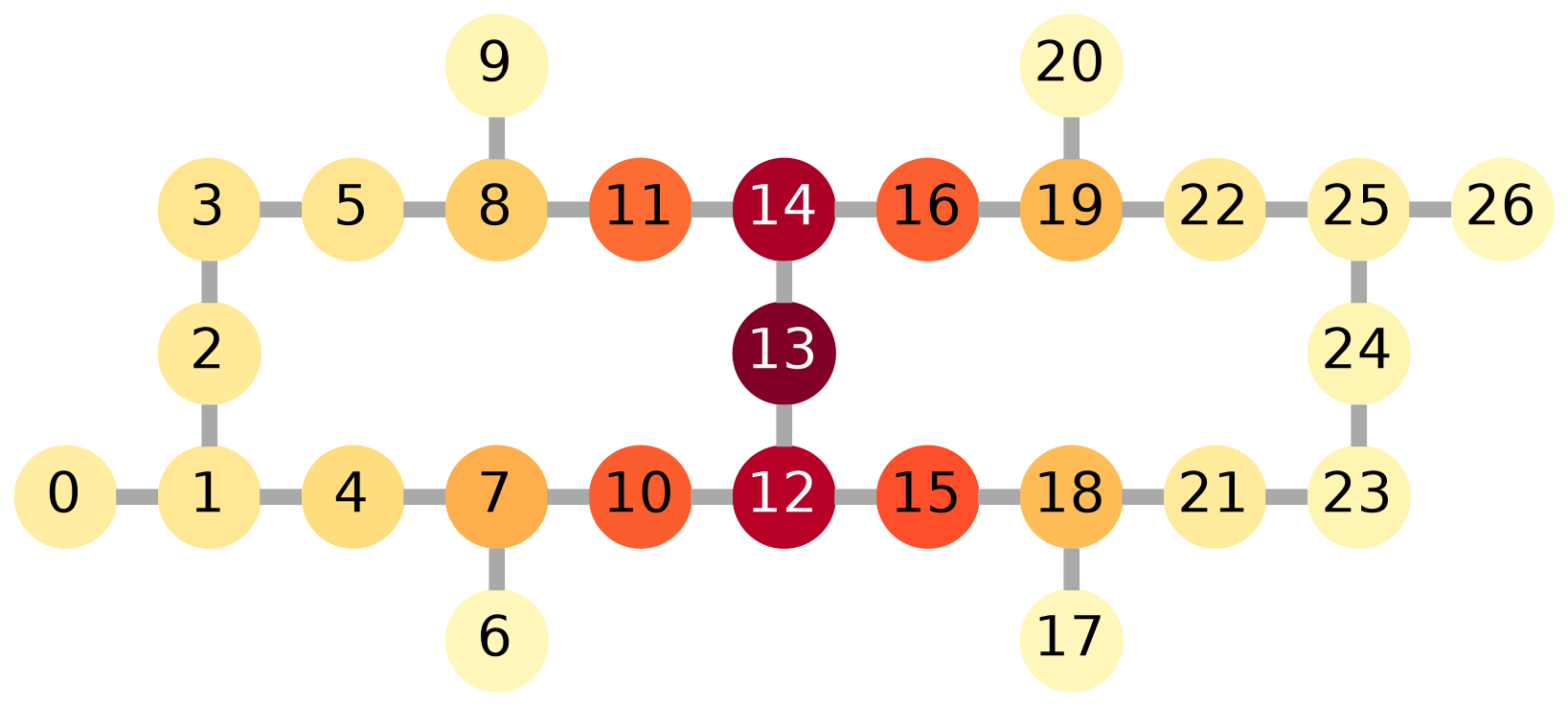}}
    \subfloat[$r=\texttt{1:1}$ and $O2$]{%
      \includegraphics[width=0.32\linewidth]{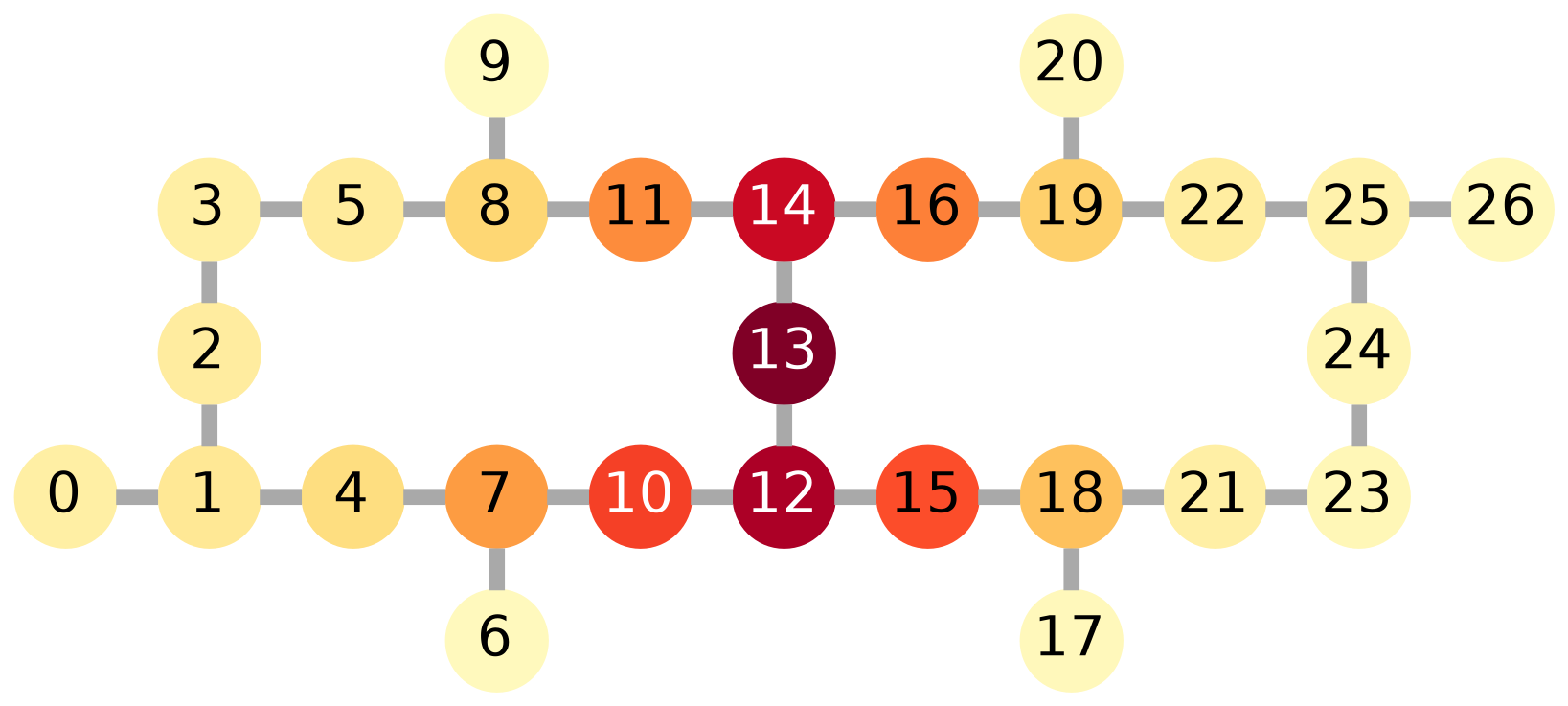}}
    \subfloat[$r=\texttt{1:4}$ and $O2$]{%
      \includegraphics[width=0.32\linewidth]{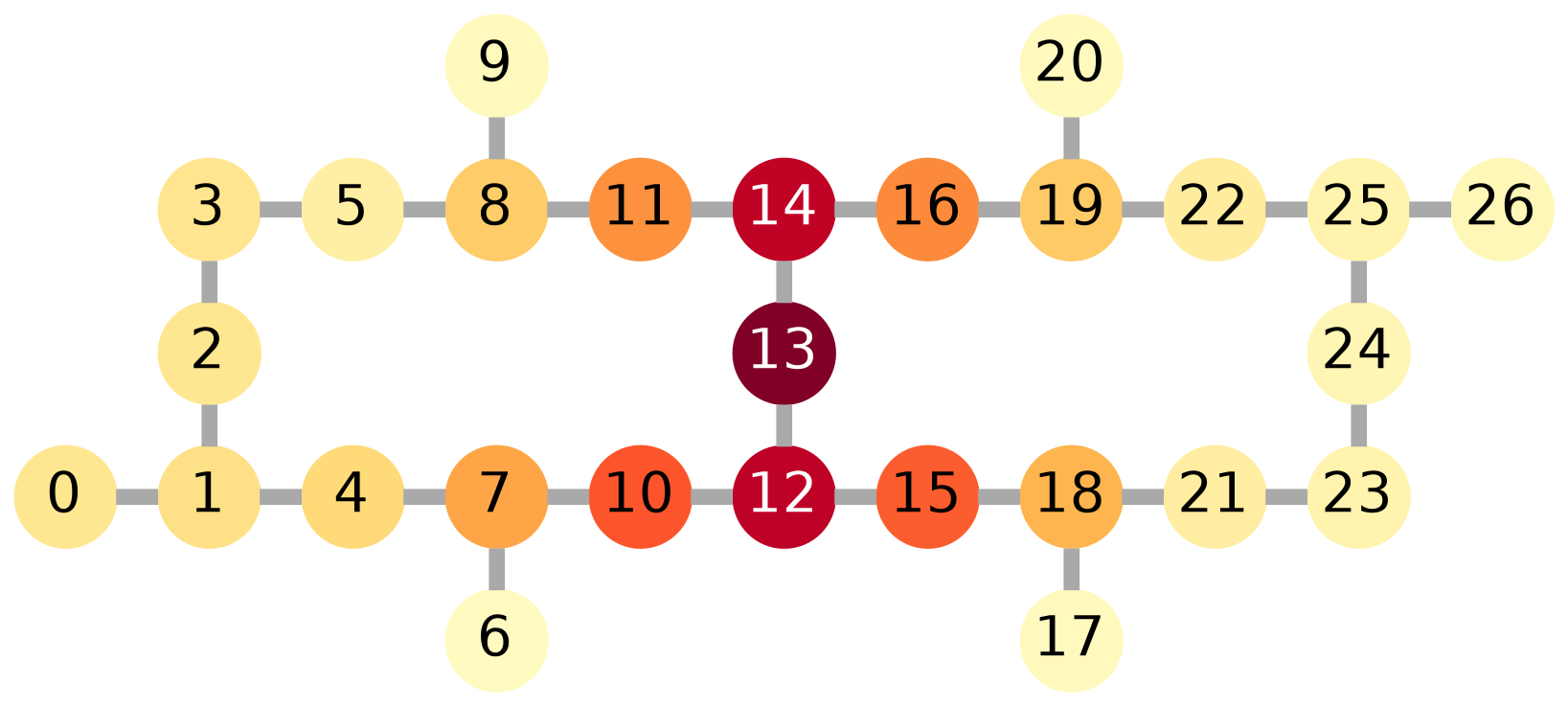}}
  \end{minipage}
  \begin{minipage}{0.11\linewidth}
    \subfloat{\includegraphics[scale=0.65]{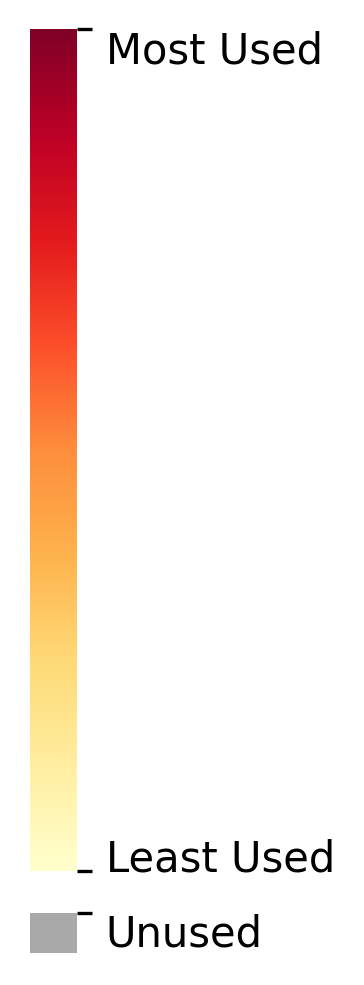}}
  \end{minipage}
  \caption{Average qubit utilization with $q=6$, $d=20$, and $L=\mathtt{sabre}$ at a sample size of $N=400$ for each image. Rows share values for $O$ and columns share values for $r$. $O1$ circuits exhibit a significant bias towards the trivial layout, which becomes more pronounced with the increasing number of $2$-qubit gates. When using $O2$, circuits homogenize with a bias towards the centermost qubit, regardless of gate distribution.}
  \label{fig:sabre:q6-d20}
\end{figure*}

Qiskit's layout and routing stages can be thought of as \textit{circuit expanding}, where qubits from the circuit are mapped in addition to those required to form optimal routing.
The optimization step effectively behaves as a \textit{circuit contractor}, potentially reducing the number of qubits used.
With the SABRE algorithm in particular, this expansion and contraction generally pulls the highest average utilization of qubits towards the qubit located closest to the center of the architecture's connectivity graph.
As seen in \autoref{fig:sabre:q6-d20}, this behavior is dependent on the coarse optimization level.
Some circuits are trivially mappable, causing a bias towards the index $0$ qubit that is correlated with higher densities of $2$-qubit gates.

\begin{figure}[!t]
  \centering
  \subfloat[$O1$]{%
    \includegraphics[width=0.8\linewidth]{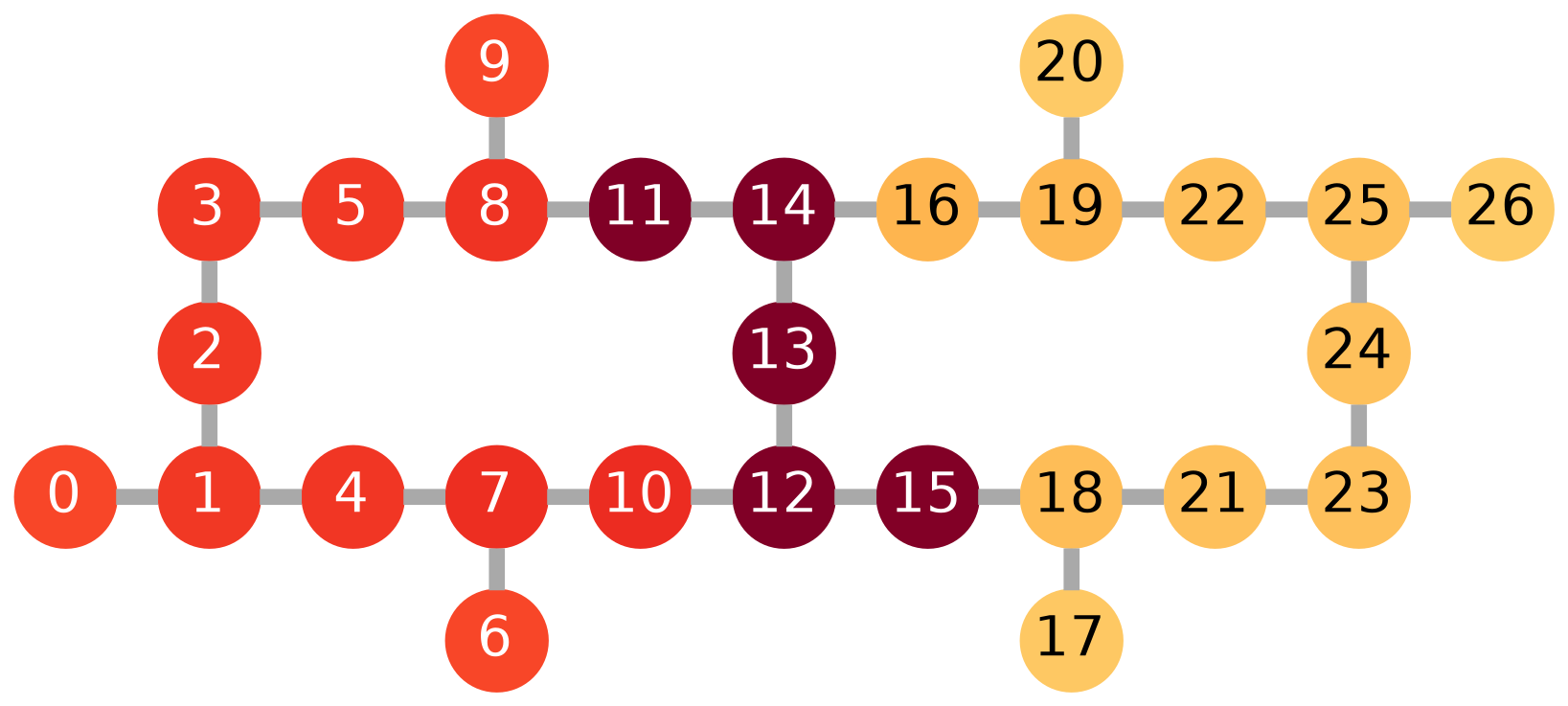}}
  \linebreak
  \subfloat[$O2$]{%
    \includegraphics[width=0.8\linewidth]{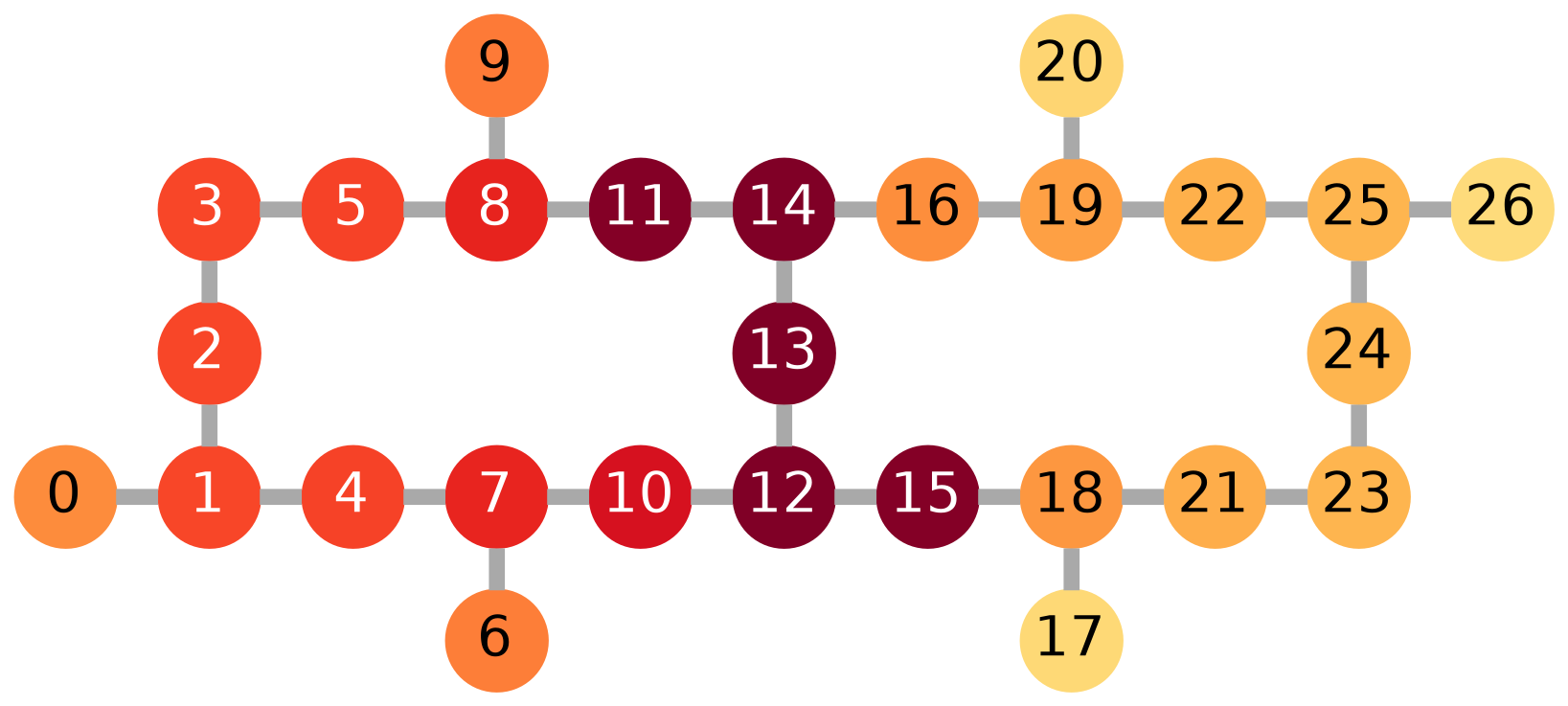}}
  \linebreak
  \includegraphics[width=0.8\linewidth]{images/colorbar_h5.png}
  \caption{Average qubit utilization for $O1$ and $O2$ using $q=16$, $d=20$, $r=\texttt{1:1}$, and $L=\mathtt{sabre}$ at a sample size of $N=400$. Shows how the bias towards the trivial mapping reemerges for circuits that utilize a majority of qubits within the architecture, regardless of optimization level.}
  \label{fig:sabre:q16-d20}
\end{figure}

When optimizing at a higher level, this trivial mapping bias gets redistributed towards the centermost qubit, index $13$, which reshapes the average qubit connectivity into a uniform gradient that decreases the further away a qubit is from the centerpoint.
This effect is invariant to the overall gate distribution and circuit depth, provided there is sufficient space to rearrange circuits within the architecture.
For high contention systems, like the one shown in \autoref{fig:sabre:q16-d20}, this behavior no longer holds, with the bias towards the trivial mapping persisting even when optimizing.
The previous contraction is still present, as evidenced by the peripheral qubits at indices $0$, $6$, $9$, $17$, and $20$ becoming less utilized.

\subsection{Limitations}
\label{sec:limitations}

\subsubsection{Architecture and Parameter Scope}

The analysis as presented is narrowly scoped on IBM's $27$-qubit Falcon R4, capping the meaningful range we can sweep other parameters.
In future work, we plan to extend this analysis to larger architectures, such as IBM's $127$-qubit Eagle R3, $133$-qubit Heron R1, and $155$-qubit Heron R2, as well as generic architectures like heavy-hex, grid, and other non-standard constructions.
Extra architectures allow much greater laterality in the parameter spans, which increases the collection and analysis times beyond what was feasible for this paper.

\subsubsection{Synthetic Workload}

Benchmarking using synthetically generated circuits is not a new concept, having been used for benchmarking hardware and software alike \cite{2025/Nation,2024/Fan,2022/Tomesh,2022/Proctor:2}.
One general downside to synthetic workloads is the strictly parameterized nature of the resulting circuit.
While we use this as a categorical distinction, the meaningfulness of those categories diminishes rapidly the further away from the collected data we infer.
To correct this, we plan to introduce several application-specific circuits utilized by quantum benchmarking suites, such as those supported by Benchpress \cite{2025/Nation}.

\subsubsection{Exclusion of Calibration Data}
\label{sec:limitations:no-calibration-data}

With NISQ computers, it is beneficial to use calibration data during layout and routing to select solutions that minimize relative error over other parameters, such as depth, which lacks correlation to runtime performance \cite{2025/Tremba}.
For the analysis presented here, we purposefully stripped all calibration data as described in \ref{sec:methodology:parameter-expansion}.
For the analysis presented here, we purposefully stripped all calibration data as described in \ref{sec:methodology:parameter-expansion}.
Analysis of how this variable affects allocation in relation to the parameters described here is ongoing.

\section{Conclusion}
\label{sec:conclusion}

In this paper, we present a novel method for determining the effect that transpiler configurations have on the average distribution of qubits for a particular architecture, with a particular emphasis on how the qubit count, circuit depth, gate distribution, layout method, and coarse optimization level change utilization patterns.
We applied this analysis to IBM's Falcon R4 architecture and found a persistent bias towards trivial mapping for lower coarse optimization levels.
While the trivial and dense methods yielded expectedly limited functionality, SABRE demonstrated an unexpected bias towards trivial mapping that persisted regardless of qubit count and gate distribution.
Instead, the coarse optimization level acted as a mitigating factor, pulling the average utilization into an approximately homogeneous gradient centered around the architecture's center point.

This initial analysis provides substantial credence to continued research into these effects, particularly in how these parameters affect larger, more capable architectures.
We plan to continue exploration of real-world workloads and parameterization, such as the effect of calibration data, to fully qualify the underlying patterns of qubit utilization and provide more broadly applicable conclusions.

\section{Acknowledgments}

This work is supported in parts by the National Science Foundation (CCF-2210963, and DGE-2113839) and gifts from Intel.

\bibliography{references}

% Generated by IEEEtranN.bst, version: 1.14 (2015/08/26)
\begin{thebibliography}{30}
\providecommand{\natexlab}[1]{#1}
\providecommand{\url}[1]{#1}
\csname url@samestyle\endcsname
\providecommand{\newblock}{\relax}
\providecommand{\bibinfo}[2]{#2}
\providecommand{\BIBentrySTDinterwordspacing}{\spaceskip=0pt\relax}
\providecommand{\BIBentryALTinterwordstretchfactor}{4}
\providecommand{\BIBentryALTinterwordspacing}{\spaceskip=\fontdimen2\font plus
\BIBentryALTinterwordstretchfactor\fontdimen3\font minus \fontdimen4\font\relax}
\providecommand{\BIBforeignlanguage}[2]{{%
\expandafter\ifx\csname l@#1\endcsname\relax
\typeout{** WARNING: IEEEtranN.bst: No hyphenation pattern has been}%
\typeout{** loaded for the language `#1'. Using the pattern for}%
\typeout{** the default language instead.}%
\else
\language=\csname l@#1\endcsname
\fi
#2}}
\providecommand{\BIBdecl}{\relax}
\BIBdecl

\bibitem[Gill et~al.(2025)Gill, Cetinkaya, Marrone, Claudino, Haunschild, Schlote, Wu, Ottaviani, Liu, Machupalli, Kaur, Arora, Liu, Farouk, Song, Uhlig, and Ramamohanarao]{2025/Buyya-Gill}
\BIBentryALTinterwordspacing
S.~S. Gill, O.~Cetinkaya, S.~Marrone, D.~Claudino, D.~Haunschild, L.~Schlote, H.~Wu, C.~Ottaviani, X.~Liu, S.~P. Machupalli, K.~Kaur, P.~Arora, J.~Liu, A.~Farouk, H.~H. Song, S.~Uhlig, and K.~Ramamohanarao, ``Chapter 2 - quantum computing: vision and challenges,'' in \emph{Quantum Computing}, R.~Buyya and S.~S. Gill, Eds.\hskip 1em plus 0.5em minus 0.4em\relax Morgan Kaufmann, 2025, pp. 19--42. [Online]. Available: \url{https://www.sciencedirect.com/science/article/pii/B9780443290961000088}
\BIBentrySTDinterwordspacing

\bibitem[Sood and Chauhan(2024)]{2024/Sood-Chauhan}
\BIBentryALTinterwordspacing
V.~Sood and R.~P. Chauhan, ``Progress and prospects of quantum computing in sustainable development: An analytical review,'' \emph{Expert Systems}, vol.~41, no.~7, p. e13389, 2024. [Online]. Available: \url{https://onlinelibrary.wiley.com/doi/abs/10.1111/exsy.13389}
\BIBentrySTDinterwordspacing

\bibitem[Coccia(2022)]{2022/Coccia}
\BIBentryALTinterwordspacing
M.~Coccia, ``Technological trajectories in quantum computing to design a quantum ecosystem for industrial change,'' \emph{Technology Analysis \& Strategic Management}, vol.~36, no.~8, pp. 1733--1748, 2022. [Online]. Available: \url{https://doi.org/10.1080/09537325.2022.2110056}
\BIBentrySTDinterwordspacing

\bibitem[AbuGhanem(2025)]{2025/AbuGhanem}
\BIBentryALTinterwordspacing
M.~AbuGhanem, ``Ibm quantum computers: evolution, performance, and future directions,'' \emph{The Journal of Supercomputing}, vol.~81, no.~5, Apr. 2025. [Online]. Available: \url{http://dx.doi.org/10.1007/s11227-025-07047-7}
\BIBentrySTDinterwordspacing

\bibitem[Castelvecchi(2023)]{2023/Castelvecchi}
\BIBentryALTinterwordspacing
D.~Castelvecchi, ``Ibm releases first-ever 1,000-qubit quantum chip,'' Dec. 2023. [Online]. Available: \url{https://www.nature.com/articles/d41586-023-03854-1}
\BIBentrySTDinterwordspacing

\bibitem[Weidman et~al.(2024)Weidman, Sajjan, Mikolas, Stewart, Pollanen, Kais, and Wilson]{2024/Weidman}
\BIBentryALTinterwordspacing
J.~D. Weidman, M.~Sajjan, C.~Mikolas, Z.~J. Stewart, J.~Pollanen, S.~Kais, and A.~K. Wilson, ``Quantum computing and chemistry,'' \emph{Cell Reports Physical Science}, vol.~5, no.~9, Sep. 2024. [Online]. Available: \url{https://doi.org/10.1016/j.xcrp.2024.102105}
\BIBentrySTDinterwordspacing

\bibitem[Sajjan et~al.(2022)Sajjan, Li, Selvarajan, Sureshbabu, Kale, Gupta, Singh, and Kais]{2022/Sajjan}
\BIBentryALTinterwordspacing
M.~Sajjan, J.~Li, R.~Selvarajan, S.~H. Sureshbabu, S.~S. Kale, R.~Gupta, V.~Singh, and S.~Kais, ``Quantum machine learning for chemistry and physics,'' \emph{Chem. Soc. Rev.}, vol.~51, pp. 6475--6573, 2022. [Online]. Available: \url{http://dx.doi.org/10.1039/D2CS00203E}
\BIBentrySTDinterwordspacing

\bibitem[Motta and Rice(2022)]{2022/Motta}
\BIBentryALTinterwordspacing
M.~Motta and J.~E. Rice, ``Emerging quantum computing algorithms for quantum chemistry,'' \emph{WIREs Computational Molecular Science}, vol.~12, no.~3, p. e1580, 2022. [Online]. Available: \url{https://wires.onlinelibrary.wiley.com/doi/abs/10.1002/wcms.1580}
\BIBentrySTDinterwordspacing

\bibitem[{Quantum Technology and Application Consortium - QUTAC} et~al.(2021){Quantum Technology and Application Consortium - QUTAC}, {Bayerstadler, Andreas}, {Becquin, Guillaume}, {Binder, Julia}, {Botter, Thierry}, {Ehm, Hans}, {Ehmer, Thomas}, {Erdmann, Marvin}, {Gaus, Norbert}, {Harbach, Philipp}, {Hess, Maximilian}, {Klepsch, Johannes}, {Leib, Martin}, {Luber, Sebastian}, {Luckow, Andre}, {Mansky, Maximilian}, {Mauerer, Wolfgang}, {Neukart, Florian}, {Niedermeier, Christoph}, {Palackal, Lilly}, {Pfeiffer, Ruben}, {Polenz, Carsten}, {Sepulveda, Johanna}, {Sievers, Tammo}, {Standen, Brian}, {Streif, Michael}, {Strohm, Thomas}, {Utschig-Utschig, Clemens}, {Volz, Daniel}, {Weiss, Horst}, and {Winter, Fabian}]{2021/Andreas}
\BIBentryALTinterwordspacing
{Quantum Technology and Application Consortium - QUTAC}, {Bayerstadler, Andreas}, {Becquin, Guillaume}, {Binder, Julia}, {Botter, Thierry}, {Ehm, Hans}, {Ehmer, Thomas}, {Erdmann, Marvin}, {Gaus, Norbert}, {Harbach, Philipp}, {Hess, Maximilian}, {Klepsch, Johannes}, {Leib, Martin}, {Luber, Sebastian}, {Luckow, Andre}, {Mansky, Maximilian}, {Mauerer, Wolfgang}, {Neukart, Florian}, {Niedermeier, Christoph}, {Palackal, Lilly}, {Pfeiffer, Ruben}, {Polenz, Carsten}, {Sepulveda, Johanna}, {Sievers, Tammo}, {Standen, Brian}, {Streif, Michael}, {Strohm, Thomas}, {Utschig-Utschig, Clemens}, {Volz, Daniel}, {Weiss, Horst}, and {Winter, Fabian}, ``Industry quantum computing applications,'' \emph{EPJ Quantum Technol.}, vol.~8, no.~1, p.~25, 2021. [Online]. Available: \url{https://doi.org/10.1140/epjqt/s40507-021-00114-x}
\BIBentrySTDinterwordspacing

\bibitem[Blunt et~al.(2022)Blunt, Camps, Crawford, Izsák, Leontica, Mirani, Moylett, Scivier, Sünderhauf, Schopf, and et~al.]{2022/Blunt}
N.~S. Blunt, J.~Camps, O.~Crawford, R.~Izsák, S.~Leontica, A.~Mirani, A.~E. Moylett, S.~A. Scivier, C.~Sünderhauf, P.~Schopf, and et~al., ``Perspective on the current state-of-the-art of quantum computing for drug discovery applications,'' \emph{Journal of Chemical Theory and Computation}, vol.~18, no.~12, pp. 7001--7023, Nov. 2022.

\bibitem[Herrmann et~al.(2023)Herrmann, Arya, Doherty, Mingare, Pillay, Preis, and Prestel]{2023/Herrmann}
\BIBentryALTinterwordspacing
N.~Herrmann, D.~Arya, M.~W. Doherty, A.~Mingare, J.~C. Pillay, F.~Preis, and S.~Prestel, ``Quantum utility - definition and assessment of a practical quantum advantage,'' in \emph{2023 IEEE International Conference on Quantum Software (QSW)}.\hskip 1em plus 0.5em minus 0.4em\relax IEEE, Jul. 2023, pp. 162--174. [Online]. Available: \url{http://dx.doi.org/10.1109/QSW59989.2023.00028}
\BIBentrySTDinterwordspacing

\bibitem[Davis(2023)]{2023/IBM/Davis}
\BIBentryALTinterwordspacing
R.~Davis, ``What is quantum utility,'' online, IBM, Nov. 2023. [Online]. Available: \url{https://www.ibm.com/quantum/blog/what-is-quantum-utlity}
\BIBentrySTDinterwordspacing

\bibitem[Kim et~al.(2023)Kim, Eddins, Anand, Wei, van~den Berg, Rosenblatt, Nayfeh, Wu, Zaletel, Temme, and Kandala]{2023/Kim}
\BIBentryALTinterwordspacing
Y.~Kim, A.~Eddins, S.~Anand, K.~X. Wei, E.~van~den Berg, S.~Rosenblatt, H.~Nayfeh, Y.~Wu, M.~Zaletel, K.~Temme, and A.~Kandala, ``Evidence for the utility of quantum computing before fault tolerance,'' \emph{Nature}, vol. 618, no. 7965, pp. 500--505, Jun. 2023. [Online]. Available: \url{https://doi.org/10.1038/s41586-023-06096-3}
\BIBentrySTDinterwordspacing

\bibitem[Beverland et~al.(2022)Beverland, Murali, Troyer, Svore, Hoefler, Kliuchnikov, Low, Soeken, Sundaram, and Vaschillo]{2022/Beverland}
\BIBentryALTinterwordspacing
M.~E. Beverland, P.~Murali, M.~Troyer, K.~M. Svore, T.~Hoefler, V.~Kliuchnikov, G.~H. Low, M.~Soeken, A.~Sundaram, and A.~Vaschillo, ``Assessing requirements to scale to practical quantum advantage,'' 2022. [Online]. Available: \url{https://arxiv.org/abs/2211.07629}
\BIBentrySTDinterwordspacing

\bibitem[Proctor et~al.(2022{\natexlab{a}})Proctor, Rudinger, Young, Nielsen, and Blume-Kohout]{2022/Proctor}
\BIBentryALTinterwordspacing
T.~Proctor, K.~Rudinger, K.~Young, E.~Nielsen, and R.~Blume-Kohout, ``Measuring the capabilities of quantum computers,'' \emph{Nature Physics}, vol.~18, no.~1, pp. 75--79, Jan. 2022. [Online]. Available: \url{https://doi.org/10.1038/s41567-021-01409-7}
\BIBentrySTDinterwordspacing

\bibitem[Murillo et~al.(2025)Murillo, Garcia-Alonso, Moguel, Barzen, Leymann, Ali, Yue, Arcaini, P\'{e}rez-Castillo, Garc\'{\i}a-Rodr\'{\i}guez~de Guzm\'{a}n, Piattini, Ruiz-Cort\'{e}s, Brogi, Zhao, Miranskyy, and Wimmer]{2025/Murillo}
\BIBentryALTinterwordspacing
J.~M. Murillo, J.~Garcia-Alonso, E.~Moguel, J.~Barzen, F.~Leymann, S.~Ali, T.~Yue, P.~Arcaini, R.~P\'{e}rez-Castillo, I.~Garc\'{\i}a-Rodr\'{\i}guez~de Guzm\'{a}n, M.~Piattini, A.~Ruiz-Cort\'{e}s, A.~Brogi, J.~Zhao, A.~Miranskyy, and M.~Wimmer, ``Quantum software engineering: Roadmap and challenges ahead,'' \emph{ACM Trans. Softw. Eng. Methodol.}, vol.~34, no.~5, May 2025. [Online]. Available: \url{https://doi.org/10.1145/3712002}
\BIBentrySTDinterwordspacing

\bibitem[Sarkar(2024)]{2024/Sarkar}
\BIBentryALTinterwordspacing
A.~Sarkar, ``Automated quantum software engineering,'' \emph{Automated Software Engineering}, vol.~31, no.~1, p.~36, Apr. 2024. [Online]. Available: \url{https://doi.org/10.1007/s10515-024-00436-x}
\BIBentrySTDinterwordspacing

\bibitem[Piattini et~al.(2021)Piattini, Serrano, Perez-Castillo, Petersen, and Hevia]{2021/Piattini}
M.~Piattini, M.~Serrano, R.~Perez-Castillo, G.~Petersen, and J.~L. Hevia, ``Toward a quantum software engineering,'' \emph{IT Professional}, vol.~23, no.~1, pp. 62--66, 2021.

\bibitem[Ito et~al.(2023)Ito, Kakimura, Kamiyama, Kobayashi, and Okamoto]{2023/Ito}
\BIBentryALTinterwordspacing
T.~Ito, N.~Kakimura, N.~Kamiyama, Y.~Kobayashi, and Y.~Okamoto, ``Algorithmic theory of qubit routing,'' 2023. [Online]. Available: \url{https://arxiv.org/abs/2305.02059}
\BIBentrySTDinterwordspacing

\bibitem[Wagner et~al.(2023)Wagner, Bärmann, Liers, and Weissenbäck]{2023/Wagner}
\BIBentryALTinterwordspacing
F.~Wagner, A.~Bärmann, F.~Liers, and M.~Weissenbäck, ``Improving quantum computation by optimized qubit routing,'' \emph{Journal of Optimization Theory and Applications}, vol. 197, no.~3, pp. 1161--1194, May 2023. [Online]. Available: \url{http://dx.doi.org/10.1007/s10957-023-02229-w}
\BIBentrySTDinterwordspacing

\bibitem[Yamanaka et~al.(2015)Yamanaka, Demaine, Ito, Kawahara, Kiyomi, Okamoto, Saitoh, Suzuki, Uchizawa, and Uno]{2015/Yamanaka}
\BIBentryALTinterwordspacing
K.~Yamanaka, E.~D. Demaine, T.~Ito, J.~Kawahara, M.~Kiyomi, Y.~Okamoto, T.~Saitoh, A.~Suzuki, K.~Uchizawa, and T.~Uno, ``Swapping labeled tokens on graphs,'' \emph{Theoretical Computer Science}, vol. 586, pp. 81--94, 2015, fun with Algorithms. [Online]. Available: \url{https://www.sciencedirect.com/science/article/pii/S0304397515001656}
\BIBentrySTDinterwordspacing

\bibitem[Javadi-Abhari et~al.(2024)Javadi-Abhari, Treinish, Krsulich, Wood, Lishman, Gacon, Martiel, Nation, Bishop, Cross, Johnson, and Gambetta]{2024/Qiskit}
A.~Javadi-Abhari, M.~Treinish, K.~Krsulich, C.~J. Wood, J.~Lishman, J.~Gacon, S.~Martiel, P.~D. Nation, L.~S. Bishop, A.~W. Cross, B.~R. Johnson, and J.~M. Gambetta, ``Quantum computing with {Q}iskit,'' 2024.

\bibitem[Jüttner and Madarasi(2018)]{2018/Juttner}
\BIBentryALTinterwordspacing
A.~Jüttner and P.~Madarasi, ``Vf2++—an improved subgraph isomorphism algorithm,'' \emph{Discrete Applied Mathematics}, vol. 242, pp. 69--81, 2018, computational Advances in Combinatorial Optimization. [Online]. Available: \url{https://www.sciencedirect.com/science/article/pii/S0166218X18300829}
\BIBentrySTDinterwordspacing

\bibitem[Zou et~al.(2024)Zou, Treinish, Hartman, Ivrii, and Lishman]{2024/Zou}
\BIBentryALTinterwordspacing
H.~Zou, M.~Treinish, K.~Hartman, A.~Ivrii, and J.~Lishman, ``Lightsabre: A lightweight and enhanced sabre algorithm,'' 2024. [Online]. Available: \url{https://arxiv.org/abs/2409.08368}
\BIBentrySTDinterwordspacing

\bibitem[Li et~al.(2019)Li, Ding, and Xie]{2019/Li}
\BIBentryALTinterwordspacing
G.~Li, Y.~Ding, and Y.~Xie, ``Tackling the qubit mapping problem for nisq-era quantum devices,'' 2019. [Online]. Available: \url{https://arxiv.org/abs/1809.02573}
\BIBentrySTDinterwordspacing

\bibitem[Nation et~al.(2025)Nation, Saki, Brandhofer, Bello, Garion, Treinish, and Javadi-Abhari]{2025/Nation}
\BIBentryALTinterwordspacing
P.~D. Nation, A.~A. Saki, S.~Brandhofer, L.~Bello, S.~Garion, M.~Treinish, and A.~Javadi-Abhari, ``Benchmarking the performance of quantum computing software for quantum circuit creation, manipulation and compilation,'' \emph{Nature Computational Science}, vol.~5, no.~5, pp. 427--435, May 2025. [Online]. Available: \url{https://doi.org/10.1038/s43588-025-00792-y}
\BIBentrySTDinterwordspacing

\bibitem[Fan et~al.(2024)Fan, Murray, Ladd, Young, and Blume-Kohout]{2024/Fan}
\BIBentryALTinterwordspacing
Y.~Fan, R.~Murray, T.~D. Ladd, K.~Young, and R.~Blume-Kohout, ``Randomized benchmarking with synthetic quantum circuits,'' 2024. [Online]. Available: \url{https://arxiv.org/abs/2412.18578}
\BIBentrySTDinterwordspacing

\bibitem[Tomesh et~al.(2022)Tomesh, Gokhale, Omole, Ravi, Smith, Viszlai, Wu, Hardavellas, Martonosi, and Chong]{2022/Tomesh}
\BIBentryALTinterwordspacing
T.~Tomesh, P.~Gokhale, V.~Omole, G.~S. Ravi, K.~N. Smith, J.~Viszlai, X.-C. Wu, N.~Hardavellas, M.~R. Martonosi, and F.~T. Chong, ``Supermarq: A scalable quantum benchmark suite,'' 2022. [Online]. Available: \url{https://arxiv.org/abs/2202.11045}
\BIBentrySTDinterwordspacing

\bibitem[Proctor et~al.(2022{\natexlab{b}})Proctor, Seritan, Rudinger, Nielsen, Blume-Kohout, and Young]{2022/Proctor:2}
\BIBentryALTinterwordspacing
T.~Proctor, S.~Seritan, K.~Rudinger, E.~Nielsen, R.~Blume-Kohout, and K.~Young, ``Scalable randomized benchmarking of quantum computers using mirror circuits,'' \emph{Phys. Rev. Lett.}, vol. 129, p. 150502, Oct. 2022. [Online]. Available: \url{https://link.aps.org/doi/10.1103/PhysRevLett.129.150502}
\BIBentrySTDinterwordspacing

\bibitem[Tremba et~al.(2025)Tremba, Hovland, and Liu]{2025/Tremba}
\BIBentryALTinterwordspacing
M.~Tremba, P.~Hovland, and J.~Liu, ``Is circuit depth accurate for comparing quantum circuit runtimes?'' 2025. [Online]. Available: \url{https://arxiv.org/abs/2505.16908}
\BIBentrySTDinterwordspacing

\end{thebibliography}
\end{document}